\documentclass{article}
\usepackage{authblk} 
\usepackage{graphicx} 
\usepackage{dcolumn}
\usepackage{bm}
\usepackage{csquotes} 
\usepackage{acronym}
\usepackage{amsmath}
\usepackage{amsfonts}
\usepackage{cleveref}
\usepackage{url}
\usepackage{amssymb}
\usepackage{caption,subcaption}
\usepackage{csquotes}
\usepackage{appendix} 
\usepackage[sorting=none,style=numeric-comp]{biblatex} 
\usepackage{pgfplots, pgfplotstable}
\pgfplotsset{compat=newest}
\pgfplotsset{minor grid style = {dashed}}

\DeclareMathOperator*{\argmin}{arg\,min}

\title{Privacy-preserving neutral atom-based quantum classifier towards real healthcare applications
}
\date{}
\author[1]{Ettore Canonici}
\author[1,2,*]{Filippo Caruso}
\affil[1]{Dept.\,of Physics and Astronomy \& European Laboratory for Non-Linear Spectroscopy (LENS), University of Florence, via G. Sansone 1, 50019 Sesto Fiorentino, Italy.}
\affil[2]{Istituto Nazionale di Ottica (INO), Consiglio Nazionale delle Ricerche (CNR), via Carrara 1, Sesto Fiorentino, 50019 Italy}
\affil[*]{Email address: filippo.caruso@unifi.it}

\begin{document}

\acrodef{nisq}[NISQ]{Noisy Intermediate Scale Quantum}
\acrodef{ml}[ML]{Machine Learning}
\acrodef{qml}[QML]{Quantum Machine Learning}
\acrodef{ai}[AI]{Artificial Intelligence}
\acrodef{qpu}[QPU]{Quantum Processing Unit}
\acrodef{svm}[SVM]{Support Vector Machine}
\acrodef{ann}[ANN]{Artificial Neural Network}
\acrodef{rnn}[RNN]{Recurrent Neural Network}
\acrodef{qubo}[QUBO]{Quadratic Unconstrained Binary Optimization}
\acrodef{qa}[QA]{Quantum Annealing}
\acrodef{qaa}[QAA]{Quantum Adiabatic Algorithm}
\acrodef{svm}[SVM]{Support Vector Machine}
\acrodef{ccf}[CCF]{Credit Card Fraud}
\acrodef{qpu}[QPU]{Quantum Power Unit}


\maketitle

\begin{abstract}
Technological advances in Artificial Intelligence (AI) and Machine Learning (ML) for the healthcare domain are rapidly arising, with a growing discussion regarding the ethical management of their development. In general, ML healthcare applications crucially require performance, interpretability of data, and respect for data privacy. The latter is an increasingly debated topic as commercial cloud computing services become more and more widespread. Recently, dedicated methods are starting to be developed aiming to protect data privacy. However, these generally result in a trade-off forcing one to balance the level of data privacy and the algorithm performance. Here, a Support Vector Machine (SVM) classifier model is proposed whose training is reformulated into a Quadratic Unconstrained Binary Optimization (QUBO) problem, and adapted to a neutral atom-based Quantum Processing Unit (QPU). Our final model does not require anonymization techniques to protect data privacy since the sensitive data are not needed to be transferred to the cloud-available QPU. Indeed, the latter is used only during the training phase, hence allowing a future concrete application in a real-world scenario. Finally, performance and scaling analyses on a publicly available breast cancer dataset are discussed, both using ideal and noisy simulations for the training process, and also successfully tested on a currently available real neutral-atom QPU.
\end{abstract}

\section{Introduction}
\ac{ml} is an area of artificial intelligence in which algorithms are able to improve their performance over time in a specific task as a result of a training process. By doing so, large amounts of data can be processed in a short time and with great efficiency. In order to do this, models need data to be trained and then to be tested. Nowadays, with the advent of digitalization, huge amounts of data are available that can be analyzed and used by models for a wide variety of purposes.
\ac{ml} has been used in various fields, such as finance \cite{Gogas2021-ad, Wasserbacher2022-kv}, the energy sector \cite{ODwyer2019-os, Szczepaniuk2023-nf}, drug discovery \cite{Vamathevan2019-rw, Vemula2023-sl}, automated anomaly detection \cite{MLanomaly2021, MLanomaly2020}, materials science \cite{Rodrigues2021-uf, Schmidt2019-wl}, and the healthcare industry \cite{Callahan2017-si, Alanazi2022-le, 8029924}.

The introduction of ML, and AI in general, into healthcare represents an opportunity to provide better services to the population. 
Potential benefits include: diagnostic efficiency, treatment personalization, increased operational efficiency, and cost reduction \cite{Hoppe2023-oz, Javaid2022-fg}.
 In healthcare applications data may come from patients who are treated daily in hospitals and at home via telemedicine. However, data contain many sensitive information that should not be shared. In general, in the use cases of \ac{ml} applied to the health sector, a primary role is given not only to the performance of the model, but also to respect for data privacy. Lastly, interpretability of the model is also a desirable requirement.. Algorithms and protocols that can satisfy all three of the above requirements are currently being researched.

Regarding privacy, numerous questions have been raised with the increasing adoption of commercial cloud computing services \cite{Sun2020-aa, Ukeje2024-vb}. Through the latter, data may be moved and processed on servers that are located across national borders, with the possibility of creating jurisdiction issues. In addition, companies may not be sufficiently encouraged to maintain privacy protections at all times if they can monetize the data or otherwise make profits from it, and legal penalties are not high enough to compensate for this behavior \cite{Murdoch2021-ic}. Information could also be stolen by others. There are attacks and breaches that cause information to be leaked. Such attacks can be directed toward the data itself or toward models \cite{privacyMLHealthcare2023}. Attacks toward data aim to steal sensitive information and in some cases are able to reconstruct the information even if anonymization techniques have been applied. Attacks toward models, on the other hand, aim to build as faithful a realization of the model used as possible. In the context of health data, data privacy violation is of greater relevance; unwanted release of the model structure is more of a problem related to companies whose business is to develop and sell models.

Currently, one of the most promising solutions for maintaining data privacy is federated learning \cite{Li2020-vt}. Although different versions of it have been formulated, such a paradigm involves dividing the dataset over several nodes while the ML model is sent in turn and trained on them in turn. 
However, currently privacy-preserving techniques often result in performance degradation compared to the classical centralized approach \cite{mammen2021federated} or still require choosing a trade-off between privacy level and performance \cite{Zhang2021-af, mo2023machine, 9069945}.
They also often make use of artificial neural networks, which are known to require a lot of data to achieve high performance.  
In any case, model interpretability is regarded as a desirable requirement.

In the near future, a further step toward privacy may be provided by quantum cryptography \cite{NielsenChuang2010, RevModPhysCriptography}. The latter, is an emerging discipline that aims to use quantum mechanical principles to provide secure channels of communication. Therefore, in the future these problems could be effectively solved by the combination of quantum cryptography and \ac{qml} models that are naturally capable of analyzing quantum data. This motivates the interest in studying and developing \ac{qml} algorithms, a relatively recent discipline that generalizes \ac{ml} in the case where at least one of the data or algorithm is quantum \cite{Biamonte2017-pk}.

The model under consideration here is a version of the \ac{svm} binary classifier whose training is reformulated into a particular class of optimization problems particularly suited to be solved by quantum processors \cite{QuboSVM2019}, namely \ac{qubo} problems \cite{QUBODefinition, QUBObook}. 
This class of problems has many applications in various fields, including graph clustering (quantum community detection problems) \cite{Negre_2020}, traffic-flow optimization \cite{Neukart2017-er}, vehicle routing problems \cite{Feld_2019}, maximum clique problems \cite{Chapuis2019-gg},  and financial portfolio management problems \cite{Mugel2021-ym}.

\ac{svm}s are a widely used class of models valued for their stability \cite{HastieTheElements2009, shalev2014understanding}, namely, the fact that small differences in the training set do not cause significant changes in the results. In general, these models are used when there is little data, but there are applications where they are used in addition to neural networks with a significant increase in performance \cite{DNNSVMkim, DNNSVMAtm, ZAREAPOOR20184}.
In addition, the fact that their training can be reformulated as a \ac{qubo} problem makes them suitable for training on a current \ac{qpu}. In particular, among all the various quantum technologies to implement \ac{qpu}s, neutral atoms are chosen here.
Indeed, although they are a young technology, they represent an attractive solution for several reasons. These include the fact that the atoms, their fundamental constituents, are naturally identical and therefore free of manufacturing errors \cite{NeutralAtoms2023}. Moreover, their scalability is related to the realization of spatially extended potentials, i.e., the possibility of creating multiple optical tweezers.

In these devices, the excited and fundamental states of the atoms are used to realize the two states of the qubit. They typically benefit from relatively long coherence times and allow work in digital mode (with unit gate sets) and analog mode. In the latter, one directly manipulates the Hamiltonian of the system, often by applying global pulses to the atoms. This greatly reduces the lifetime of quantum protocols and the number of operations performed on the qubits, which is particularly important in \ac{nisq} devices \cite{preskill2018quantum}.
However, there are studies on how to reduce the effects of noise even in neutral atom devices to try to increase their applicability and usefulness \cite{Bluvstein2023-li,Canonici2024-fw}, already obtaining at least 48 logical qubits.

In neutral atom devices, \ac{qubo} problems can be solved by \ac{qaa} \cite{Albash_2018}, a process in which a quantum Hamiltonian evolves very slowly from one of easy realisation into that of the desired problem. The state of the system, in turn, changes from the initial fundamental state to that of the desired Hamiltonian, resulting in the solution of the problem.
Potentially, the high connectivity and flexibility of the atom topology and the Rydberg mechanism could make them more advantageous than other solutions.

Among the advantages of such a model is that it does not require the data to be moved to the cloud. In fact, if the \ac{qpu} is available via the cloud, only the coordinates of the atoms and the intensity and duration values of the laser pulses will need to be provided. This seems to provide a paradigm-model that preserves data privacy and is relatively interpretable. In fact, \ac{svm} models are generally believed to be more interpretable than neural networks and quantum variational circuits. 
The model presented here is based on analog quantum computing, which unlike digital quantum gates-based quantum computing allows for fewer operations to be performed on the qubits and circumvents the problems arising from the circuit depth lengthening caused by encoding given that in our case it is done through the topology of atoms. Moreover, in our case the \ac{qpu} is only for training the model. Since the testing phase takes place locally and not on the cloud, there is a reduction in the use of the \ac{qpu} and no data is released via the cloud.

Unlike \cite{Watkins2023-qa} our approach, requires that only the training exploits a \ac{qpu} and thus is applicable on current NISQ devices.
To ensure that even the last of the 3 requirements of AI for applications in the healthcare domain, we verify that the performance of the model is adequate by testing on a simple healthcare dataset.
We propose an analysis performed with various versions of the QUBO SVM model and compare the results with the main models in the literature. As for the QUBO SVM model, it was implemented and simulated using the \emph{Python} library \emph{Pulser} \cite{Silverio2022Pulser}, which allows ideal and noisy simulations.
\emph{Pulser} is developed by \emph{Pasqal} \cite{pasqal}, a company that builds quantum processing units based on neutral atoms capable of using up to 100 qubits. Through \emph{Pulser}, it is possible to run simulations on both real machines and an integrated simulator.
Therefore, what we want to show is that this solution offers an excellent combination of performance, and potentially allows it to be used in contexts of sensitive data and where the predictive model needs to be explainable.

\section{Methods}

\subsection{QUBO problem}

Within the branch of mathematical optimisation, there are a number of problems that lend themselves to being solved using quantum processors. These combinatorial optimisation problems, known as \ac{qubo} problems, are characterised by peculiar connections with Ising models and for this reason possible connections with physics are studied in order to be able to find solutions more efficiently. In general, they are NP-hard problems, i.e. there are currently no algorithms capable of solving them in polynomial time. Indeed, in some cases they cannot even be solved in polynomial time.
For this reason, many methodologies have been explored for their approximate solution, including quantum computing. In fact, in some particular cases quantum processors have proved superior in their approximate solution.
From a mathematical point of view, an \ac{qubo} problem is defined by the following quadratic cost function:

\begin{equation}\label{eq: QUBO def}
\begin{aligned}
E &= a^{\top} Q a \\
  &= \sum_{i,j=1}^{n} a_i Q_{ij} a_j \\
  &= \sum_{i=1}^{n} a_i Q_{ii} + \sum_{i \neq j}^{n} a_i Q_{ij} a_j
\end{aligned},
\end{equation}
with $Q_{ij} \in \mathbb{R} \,\, \forall \, i,j$.
In QUBO problems the variables $a$ are binary, i.e. $a \in\mathbb\{0,1\}^{n} = \mathbb{B}^{n}$. The solution of the problem
\begin{equation}
    a^* = \argmin_{a \in\mathbb{B}^{n}} E,
\end{equation}
is the variable that minimises the cost function $E$

\subsection{Support Vector Machines}

Among the various machine learning tasks, one of the most popular is binary classification. A parametric model is trained via supervised learning to assign labels to two different categories.
Among the various models, some of the most popular are the so-called Support Vector Machines (SVMs). Requiring much less data than neural networks, SVMs are still valid choices in the case of small to medium-sized datasets, offering good performance, robustness and explainability. Moreover, through kernels, they can be used on linearly and non-linearly separable datasets.
Let us now consider a dataset
\begin{equation}
    D = \{ (x_n, y_n) : x_n \in\mathbb{R}^{d}, y_n = \pm 1 \}_{n=0, \dots, M-1}, 
\end{equation}
with $x_n$ sample (also known as feature vector) and $y_n \in \{-1, 1\}$ associated target. The samples of the class marked $y_n = 1 \,\,(-1)$ are called \textquote{positives} (\textquote{negatives}).

The training problem of the SVM model is as follows:
\begin{equation}\label{svm training}
    \text{minimize }  [ \frac{1}{2} \sum_{n,m} \alpha_n \alpha_m y_n y_m k(x_n, x_m) - \sum_{n} \alpha_n ]
\end{equation}
\begin{equation}\label{first constraint}
    \text{subject to   } 0 \leq \alpha_n \leq C,
\end{equation}
\begin{equation}\label{second constraint}
    \text{and   } \sum_n \alpha_n y_n = 0,
\end{equation}
where $C$ is a regularisation constant and $k(\cdot)$ is called a kernel function.

The coefficients $\alpha_n$ define a hyperplane that divides $\mathbb{R}^{d}$ into two regions, one for each class. At this point, given a boundary decion corresponding to a certain $\alpha$, a generic sample $x_j$ can be assigned to one of the two classes via:
\begin{equation}
    \hat{y_j} = sgn(f(x_j)),
\end{equation}
dove
\begin{equation}\label{bias}
   b = \frac{ \sum_n \alpha_n (C -  \alpha_n)[y_n - \sum_m \alpha_m y_m k(x_n, x_m)]}
   { \sum_n \alpha_n (C -  \alpha_n)}.
\end{equation}
is the decision function.It represents the marked distance of the sample from the decision boundary. Furthermore, the bias $b$ is defined by the following:
\begin{equation}\label{bias2}
   b = \frac{ \sum_n \alpha_n (C -  \alpha_n)[y_n - \sum_m \alpha_m y_m k(x_n, x_m)]}
   { \sum_n \alpha_n (C -  \alpha_n)}.
\end{equation}

Furthermore, having introduced the QUBO problems and the SVM model, it is possible to rewrite Eq.\ref{svm training} in QUBO form. By doing so, the training problem of an SVM model is related to the minimisation of a cost function of the type Eq.\ref{eq: QUBO def}.
To do this, one must first encode the variables of the problem in the form of binary variables, since analog quantum processors can only produce binary solutions.
The encoding is done via the following variables:
\begin{equation}\label{encoding}
    \alpha_n = \sum_{k=0}^{K-1} B^k a_{Kn+k}.
\end{equation}
In this case we fix a basis for encoding, $B$, and to encode the variable $\alpha_n$ we use $K$ binary variables  $ a_{Kn+k} \in \{ 0,1\}$.
At this point, making use of the multiplier $\xi$, it is possible to rewrite Eq. \ref{eq: QUBO def} in the following form:
\begin{equation}\label{qubo energy}
    E = \sum_{n,m=0}^{N-1} \sum_{k,j=0}^{K-1} 
     a_{Kn+k} \Tilde{Q}_{Kn+k, Km+j} a_{Km+j}
\end{equation}
with 
\begin{equation}\label{qubo matrix}
    \Tilde{Q}_{Kn+k, Km+j} = \frac{1}{2}B^{k+j} y_n y_m (k(x_n, x_m) + \xi) - \delta_{nm}\delta_{kj}B^k.
\end{equation}
At this point, it can be observed that the QUBO matrix in \ref{qubo matrix}, has dimension $KN \times KN$, where $N$ is the dimension of the training set.
For a more in-depth derivation, see \cite{Canonici2024QUBO}.

\subsection{Implementation on a neutral atom QPU}

\begin{figure}[h!]
    \centering
\includegraphics[width=1.\textwidth]{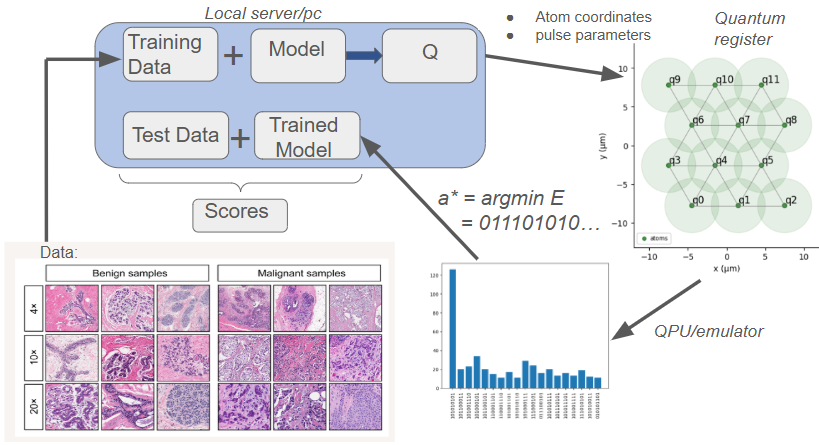}
\caption{Figure containing a schematic representation of the entire approach. The data is never transmitted via the cloud to the QPU, preventing its unwanted disclosure.}
\label{fig:full_pipeline}
\end{figure}

\begin{figure}[h!]
    \centering
\includegraphics[width=1.\textwidth]{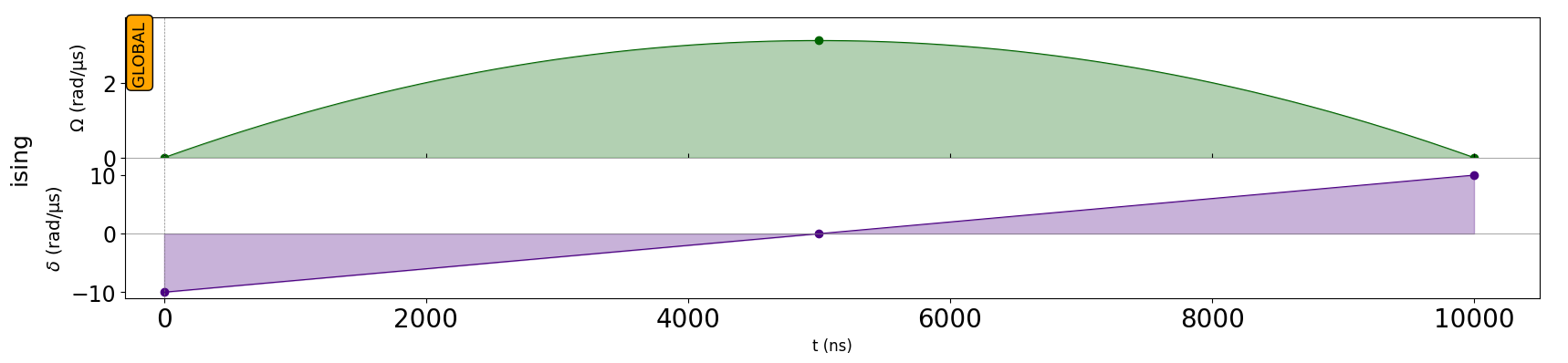}
\caption{Pulse sequence employed for the training of the QUBO SVM model. The waveforms of the Rabi frequency and detuning are depicted in green and purple, respectively. The maximum height of the former is ascertained in \ref{eq:omega max}.}
\label{QAA_pulse}
\end{figure}

After the model has been introduced, it is possible to describe its implementation on a real neutral-atom quantum processor. A graphical representation of the entire process is in Fig.\ref{fig:full_pipeline}.
Furthermore, as explained in \cite{Canonici2024QUBO}, it is possible to divide the implementation of the model into three phases: encoding, training and testing of the model. Only the second phase requires the use of the QPU, while the other two exploit classical hardware.
Before doing so, let us recall that a system of interacting neutral atoms is described by the following Hamiltonian:
\begin{equation}\label{eq: rydberg hamiltonian}
    H = \frac{\hslash}{2} \sum_{i} \Omega(t)\sigma_{i}^{x} - \hslash  \sum_{i} \delta(t) \hat{n}_i + \sum_i \sum_{i<j} U_{ij} \hat{n}_i \hat{n}_j,
\end{equation}
where $\Omega$ and $\delta$ are the Rabi frequency and detuning, respectively, while $\sigma^k, \, k \, \in \, \{x,y,z\}$ are the Pauli matrices and $n_i = (1 + \sigma_{i}^{z})/2$.
By positing $\Omega = 0$ in Eq. \ref{eq: rydberg hamiltonian}, one realises its similarity to Ising's model and Eq. \ref{eq: QUBO def}. In fact, it is possible to show that these are similar provided a symmetry operation is performed:
\begin{equation}\label{eq:QUBO simmetric}
Q_{ij} = 
\begin{cases} 
    \frac{Q_{ij} + Q_{ji}}{2} & \text{if } i \neq j \\
    Q_{ii} & \text{otherwise}
\end{cases}.
\end{equation}
As already described in \cite{Canonici2024QUBO}, $NK$ atoms are needed to encode the $N$ training samples. In this case, $K=2$ was chosen. At this point, using the COBYLA optimiser \cite{cobyla1992}, we search for the distribution of atoms in space such that the Hamiltonian of the system is as close as possible to the $\Tilde{Q}$ matrix, containing the information about the training data of the SVM model rewritten in QUBO form. To do this, the physical constraints of quantum hardware must also be imposed, i.e. a maximum of 25 atoms, atoms with a minimum relative distance greater than 5 $\mu m$ and a maximum distance from the centre of the quantum register of 35 $\mu m$. Furthermore, for the runs on the QPU we had to impose the additional constraint of having the atoms on the vertices of a register with a triangular grid geometry. What has just been said is part of the encoding step of the problem, the training of the model on the other hand consists of running the adiabatic algorithm to find the state corresponding to the minimum of the cost function. To perform the adiabatic evolution, we used both the Pulser emulator and the real QPU. The basic idea behind adiabatic evolution is to start in the ground state of an easy-to-prepare Hamiltonian and then slowly change it to that of the desired problem (which is obtained exactly for $\Omega = 0$). If the evolution has been sufficiently smooth, the ground state of the final Hamiltonian encoding the problem is obtained. In practice, a sequence of pulses implementing adiabatic evolution is shown in Fig.\ref{QAA_pulse}.

As done in \cite{Canonici2024QUBO}, the sequence of pulses for detuning is a linear ramp from -10 to + 10 $rad/\mu s$ that lasts for a time $\tau = 10 \, \mu s$. As far as Rabi's frequency is concerned, the pulse has the form of a bell with a maximum value of
\begin{equation}\label{eq:omega max}
    \begin{cases}
    median(\Tilde{Q}),& \text{if }  median(\Tilde{Q})< \Omega_{max}\\
    \Omega_{max} = 15.71 \,\, rad/\mu s,              & \text{otherwise},
\end{cases}
\end{equation}
and duration $\tau = 10 \, \mu s$. 
However, on the QPU the durations were reduced to $\tau = 4 \, \mu s$.

The simulation is finally repeated $N_{shots}$ times. This yields a distribution of observed states and their frequencies, which can be normalised to obtain probabilities. Each state obtained represents a possible decision boundary defined by $\alpha_n$. The state with the highest probability is the solution of the QUBO problem, i.e. the SVM model that minimises errors.
At this point, to test the model, it is sufficient to use a particular value of $\alpha_n$ associated with the maximum likelihood quantum state and use the model on the test data and compute the desired metrics.
However, with a single training of $N_{shots}$, up to $2^{N \times K}$ different models can be obtained, one for each state of the computational basis of an $N \times K$ qubits system. Therefore, ensemble learning schemes can be implemented. It is based on the assumption that the combined predictions of several models are generally better than those of a single model.
There are various versions of ensemble models, including average voting and stacking. In the former paradigm, the final prediction is the average of the predictions of the individual models. In contrast, in stacking, the predictions of a group of models are used to train a meta-model, which will serve as the second level. In any case, further details can be found in \cite{Canonici2024QUBO, HastieTheElements2009}.

Since, as already mentioned, we obtain up to $2^{N \times K}$ models with a single training, we can naturally use them for a quantum average voting strategy. For this reason, we start with the model associated with the state with the highest probability and then consider the models associated with gradually decreasing probabilities. By doing so, it is possible to optimise the number of models in such a way as to maximise a particular metric. This is possible by specifying a validation set.
It is also possible to use the QUBO SVM model as a meta-model in ensemble learning schemes, where the basic models are classical.
In addition, it is possible to observe how only the coordinates of the atoms and the pulse sequence need to be provided to the QPU in order to perform the quantum simulations required for training. This avoids the disclosure of sensitive data and information they contain, which is particularly crucial in certain applications where data privacy is a prerequisite.

\subsection{Dataset and Metrics}

The dataset that was chosen to test the model is a famous health dataset also present on scikit-learn \cite{scikit-learn}. It contains data from breast cancer patients. It has a total of 569 samples divided into two classes: benign and malignant tumors, respectively 357 and 212. Each sample has 30 features, containing information regarding the tumor. Such dataset is widely used to test binary classifiers.

The data are standardized, then divided into two parts. $60\%$ percent of the data was used to construct a dataset from which the training set and validation set were extracted. The training contains 6 training samples. The remaining 40$\%$ is kept apart and used later as the test set. 
The entire process is repeated 10 times, thus selecting 10 different splits. In the end, 10 values are then obtained for the main selected metrics. To obtain an estimate, the metrics are provided in the form of the mean value and standard deviation calculated on the 10 experiments. In addition, the procedure is repeated by increasing the number of training samples to 7 and 8 (using 14 and 16 atoms respectively).

In this case, the dataset being balanced, the metrics considered are accuracy and F1 score.
Accuracy, quite simply, is defined as the ratio between the number of correct predictions and the total number of samples analysed.
While accuracy can provide a general idea of how well the model works, it has some limitations. In particular, in cases where the dataset is highly unbalanced the model prefers to learning to recognize only samples in the majority class well. In extreme cases the model always assigns the sample in the majority class to maximize accuracy. In reality this results in low predictive power and poor generalization abilities of the model. Finally, by maximizing accuracy there is no control over the minimization of false positives and false negatives.
To solve these problems, metrics such as F1 score can be used. It is defined as the harmonic mean of precision and recall, two metrics widely used in classification. Recall (sensitivity) is the ratio of correctly predicted positive observations to all actual positives while precision (specificity) is the ratio of correctly predicted positive observations to the total predicted positives. 
Therefore by maximizing F1 score one tries to maximize precision and recall simultaneously, therefore one tries to minimize false positives and false negatives simultaneously. It is defined as:
\begin{equation}
\begin{split}
    \text{F1 score} =  2 \times \frac{precision \times recall}{precision + recall} \\ = \frac{2 TP} {2 TP + FP + FN}.
\end{split}
\end{equation}
where $TP$, $FP$ and $FN$ are used to indicate true positives, false positives and false negatives respectively.
F1 score varies between 0 and 1, with 1 being the best obtainable value. This metric is among those that can be used in unbalanced class contexts.
While accuracy is a good starting point, it does not give the whole picture. Therefore F1 score provides a more complete overview of model performance.

\section{Results and Discussion}  
\begin{figure}
\begin{subfigure}{.5\textwidth}
  \centering
  \includegraphics[width=0.95\linewidth, height=4.5cm]{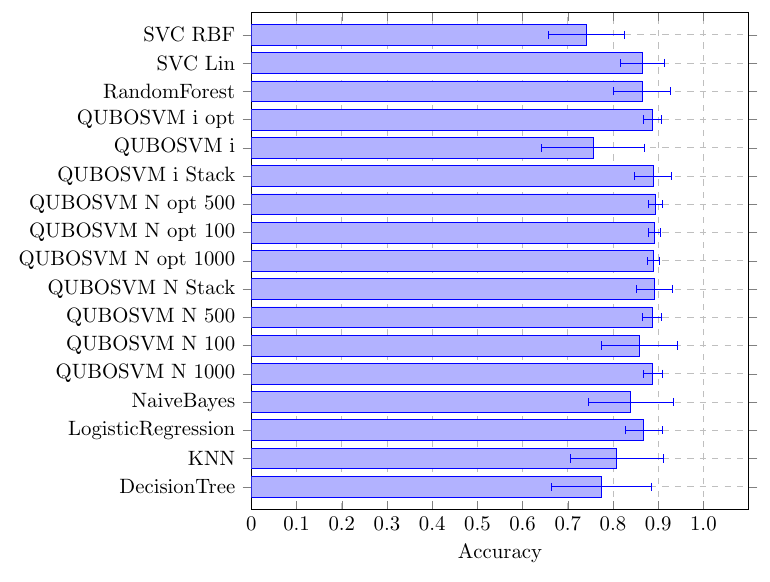}  
  \caption{6 training samples, 12 qubits simulations.}
  \label{fig:Acc6}
\end{subfigure}
\begin{subfigure}{.5\textwidth}
  \centering
  \includegraphics[width=0.95\linewidth, height=4.5cm]{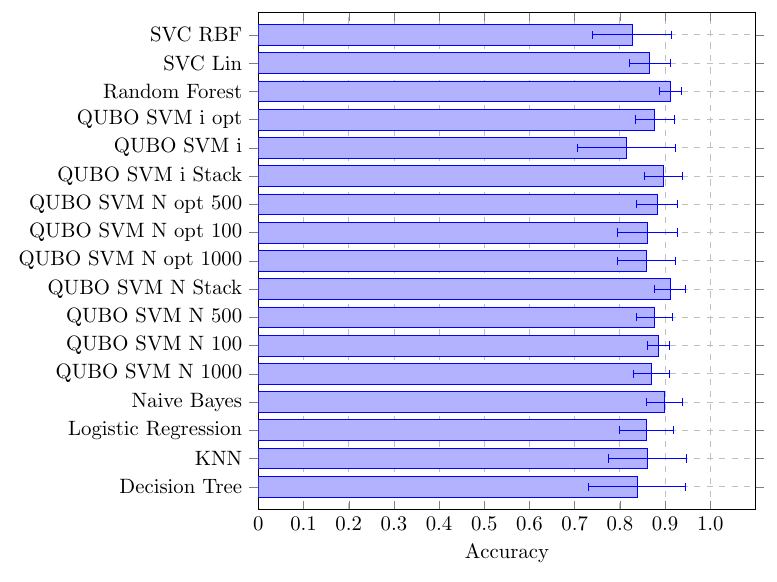}  
  \caption{7 training samples, 14 qubits simulations.}
  \label{fig:Acc7}
\end{subfigure}
\newline
\begin{subfigure}{.5\textwidth}
  \centering
  \includegraphics[width=0.95\linewidth, height=4.5cm]{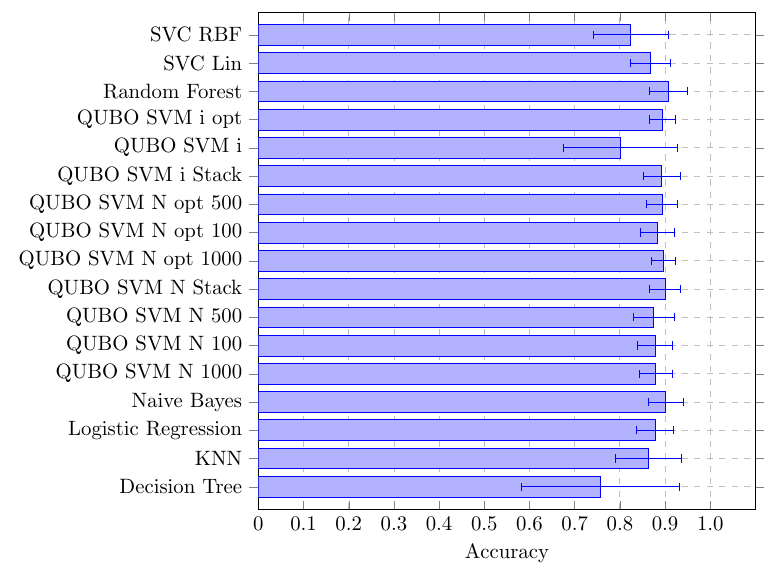}  
  \caption{8 training samples, 16 qubits simulations.}
  \label{fig:Acc8}
\end{subfigure}
\caption{Comparison of the accuracy of different models using datasets of 6 (a), 7 (b) and 8 (c) samples. In each subfigure, the results are shown in the form of average values and error bars (standard deviation).}
\label{fig:Accuracy}
\end{figure}

\begin{figure}
\begin{subfigure}{.5\textwidth}
  \centering
  \includegraphics[width=0.95\linewidth, height=4.5cm]{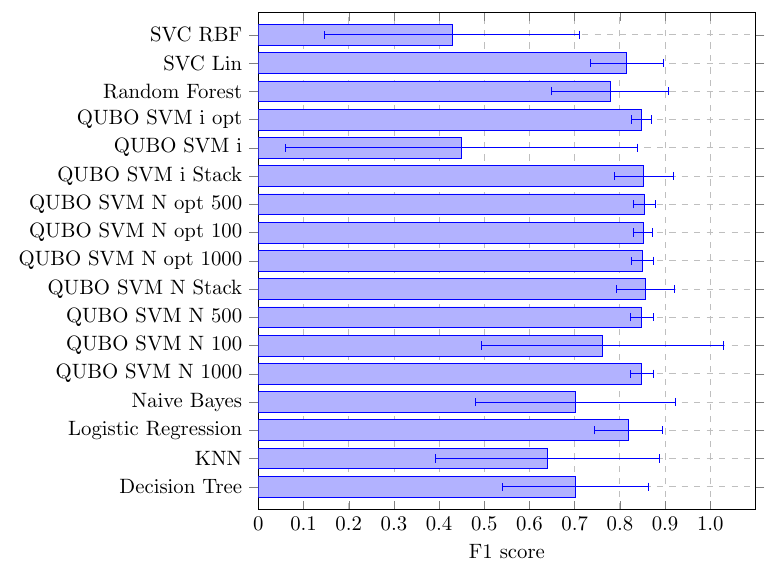}  
  \caption{6 training samples, 12 qubits simulations.}
  \label{fig:f16}
\end{subfigure}
\begin{subfigure}{.5\textwidth}
  \centering
  \includegraphics[width=0.95\linewidth, height=4.5cm]{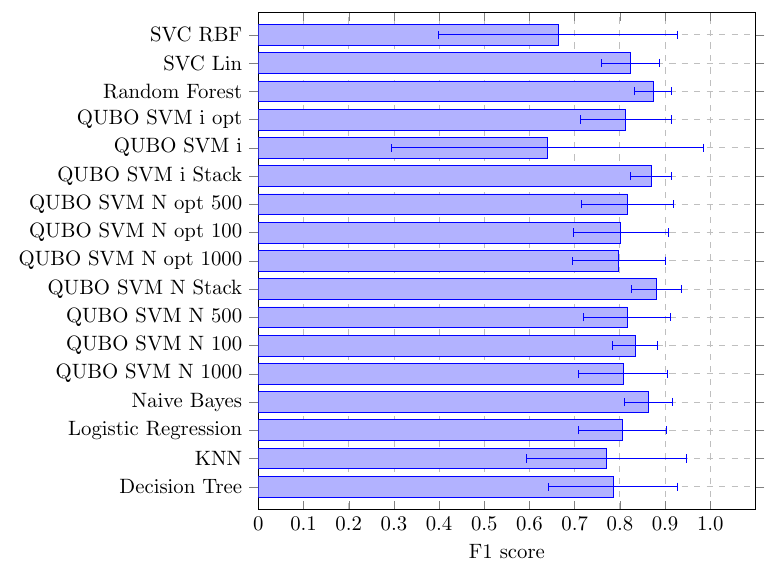}  
  \caption{7 training samples, 14 qubits simulations.}
  \label{fig:f17}
\end{subfigure}
\newline
\begin{subfigure}{.5\textwidth}
  \centering
  \includegraphics[width=0.95\linewidth, height=4.5cm]{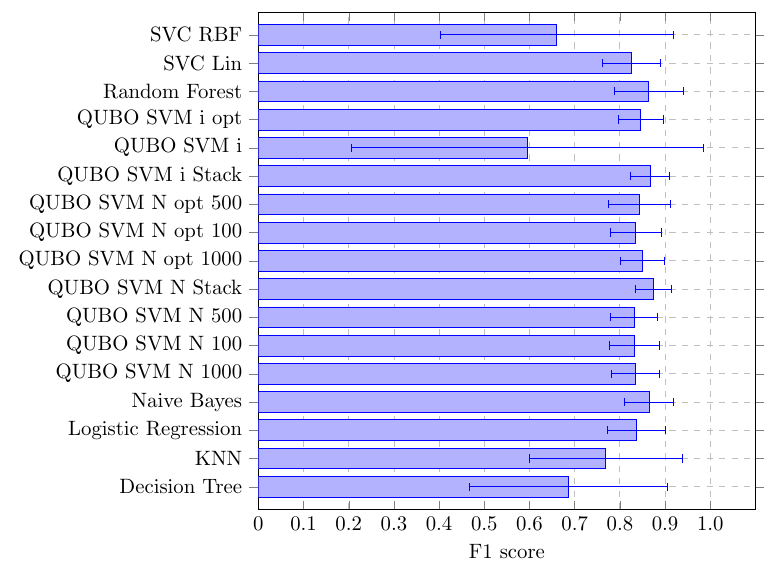}  
  \caption{8 training samples, 16 qubits simulations.}
  \label{fig:f18}
\end{subfigure}
\begin{subfigure}{.5\textwidth}
  \centering
\end{subfigure}
\caption{Comparative analysis of the F1 scores of all the different models considered, using datasets of 6 (a), 7 (b) and 8 (c) samples. In each subfigure, the results are shown in the form of average values and error bars (standard deviation).}
\label{fig:F1}
\end{figure}

This section describes numerical experiments, ideal and in the presence of noise, related to the training of QUBO SVM models.
An exhaustive list of all models used can be found in Appendix \ref{appendix:Models}.
Performance analysis including scaling in the size of the training set is also performed. The results are compared with results obtained by classical ML models and are shown in Figs.\ref{fig:Accuracy} and \ref{fig:F1}.

Fig.\ref{fig:Accuracy} and Fig.\ref{fig:F1} show, respectively, the accuracy and F1 scores of various models (classical and trained with emulator) at varying training set sizes, using 6, 7 and 8 samples, respectively.
It can be seen that the QUBO SVM models seem to perform well regardless of the number of runs of the algporhythm used for training, 100, 500 and 1000.
This is interesting, as being able to use only 100 per training means using quantum hardware in a reduced way, with potential savings in time and costs associated with computing services. 
Furthermore, as also shown in \cite{Canonici2024QUBO}, the models trained with and without noise (denoted by \textquote{N} and \textquote{i}, respectively) in simulations behave similarly, showing some robustness to noise without generally showing excessive drops in performance.
Finally, the model in stack configuration seems to offer the best mix of perfrormance goodness and error bar size.

\begin{figure}[ht]
\begin{subfigure}{.5\textwidth}
  \centering
  \includegraphics[width=.8\linewidth]{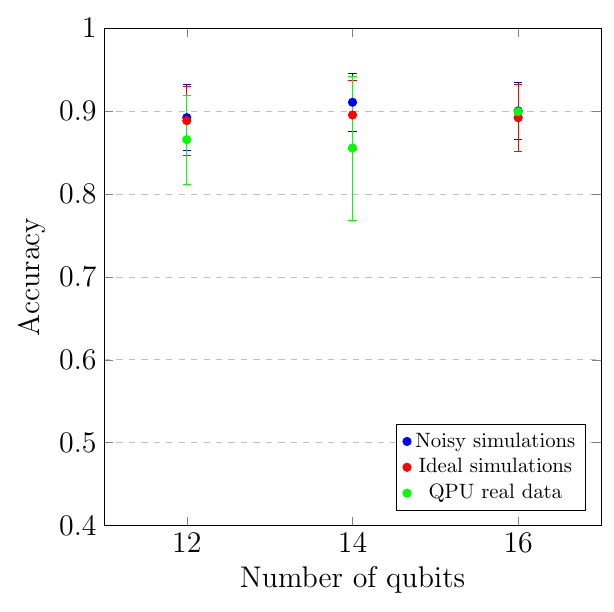}  
  \caption{}
  \label{fig:QPU accuracy}
\end{subfigure}
\begin{subfigure}{.5\textwidth}
  \centering
  \includegraphics[width=.8\linewidth]{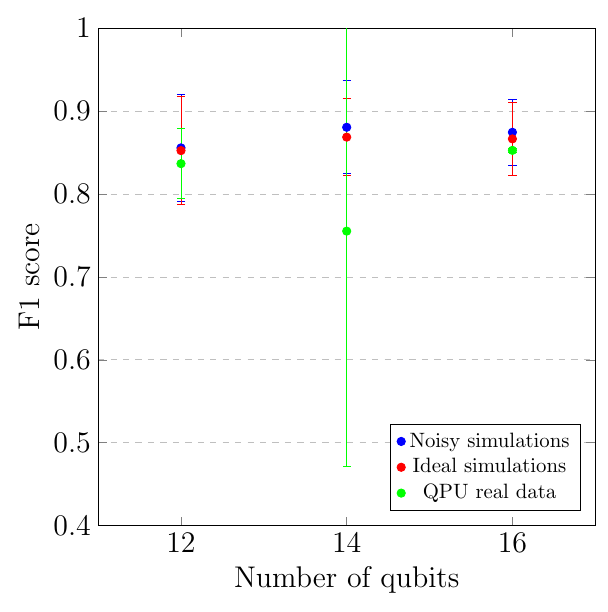}  
  \caption{}
  \label{fig:QPU F1 score}
\end{subfigure}
\caption{Accuracy (a) and F1 score (b) scaling with the number of qubits (atoms).}
\label{fig:QPU}
\end{figure}

We now consider the results obtained by training the models not on a quantum emulator, but directly on a Pasqal QPU prototype. It allows the use of atomic registers containing a maximum of 25 Rubidium atoms trapped by optical tweezers.
We therefore perform a comparison of the performance of the QUBO SVM model trained with an emulator (ideal and non-ideal) and the real QPU. The results are shown in Fig.\ref{fig:QPU} and contain the model accuracies as a function of the number of atoms used (Fig.\ref{fig:QPU accuracy}) and the F1 score (Fig.\ref{fig:QPU F1 score}). With the exception of one point where the QPU-trained version scores worse performance and significantly larger error bars, probably due to some error during the quantum simulation, there seems to be no significant difference. This is important because the performance on the real harwadre is always lower due to quantum noise, especially in models implemented through quantum circuits and based on quantum gates. Furthermore, it is interesting to observe that the performance does not decrease as the number of atoms increases, since in general an increase in the size of the quantum system is associated with an increase in the possible sources of error and thus a lower overall efficiency of the algorithms.


\section{Conclusion}

We have introduced an SVM model whose training is recastt in the form of a QUBO problem. It can be naturally implemented on neutral atom devices. This model is based on analog processing, i.e. only global operations are performed on the atoms.

In particular, the considered model is a linear classifier, and can also be applied to non-linearly separable data via feature engineering and via kernel.
The model was trained and tested on a famous breast cancer detection health dataset.
Several variants of the model were proposed and tested, both under ideal conditions and in the presence of noise. For each version, 10 training sates were run on 10 splits of the dataset in order to provide more robust performance estimates. Finally, 3 different scenarios were considered in which the training sates had 6, 7 and 8 sampels respectively. This required the use of quantum registers of 12, 14 and 16 atoms respectively.

Furthermore, comparisons are made between the QUBO SVM models and some of the main classical binary classifiers. In general, the performances obtained are encouraging and seem to be on par, if not slightly better, than the analysed classical models, probably also due to the proposed ensemble formulation.
The models also show a certain robustness to noise, also assessed through simulations on a real QPU with neutral atoms, which makes the presented pipeline already appliable in real scenarios, even those where data is sensitive. In fact, as already expressed, only information concerning the quantum register topology and pulse sequence is transmitted to the QPU.
Since our protocol involves using the QPU only in the training phase, it is suitable for application to real-world scenarios, where costs must be kept low.
The fact that the model is SVM also ensures a certain interpretability to the results obtained, which is useful in decision-making contexts.

Finally, we think we can scale to larger training sets through the use of atomic QPUs with a few hundred atoms already available via the cloud, and through the application of techniques such as branch-and-bound. Such techniques break down an optimisation problem into smaller parts that are solved separately. Subsequently, the global solutions are assembled to obtain the solution of the starting problem. For example, using the technique described in \cite{vandelli2025parallel} together with QPUs of more than 200 atoms, we count on being able to scale to training sets of several hundreds, or a few thousand, samples.
The fact that the model is SVM also ensures a certain interpretability to the results obtained, which is useful in decision-making contexts.

In the future, we will study the performance obtained by using larger QPUs and branch-and-bound techniques as just described. This will allow us to test these types of models in increasingly complex scenarios.

\section{Acknowledgements}
This work has been financially supported by the European High-Performance Computing Joint Undertaking (JU) under grant agreement No. 101018180 – HPCQS, by the PNRR MUR project PE0000023 – NQSTI, by the European Commission’s Horizon Europe Framework Programme under the Research and Innovation Action GA No. 101070546 – MUQUABIS, and by the MUR Progetti di Ricerca di Rilevante Interesse Nazionale (PRIN) Bando 2022 – Project No. 20227HSE83 – ThAI-MIA, funded by the European Union – Next Generation EU.
Finally, the authors thank Pasqal for running the simulations on their QPU and for the support shown during the development phase.


%
\printbibliography

\appendix\section{Models}\label{appendix:Models}

All used models are listed here, including all versions of the QUBO SVM models:
\begin{itemize}
    \item K-nearest neighbors (KNN)
    \item Random Forest
    \item Decision Trees
    \item Naive Bayes
    \item Logistic Regression
    \item SVM with linear kernel
    \item SVM with rbf kernel
    \item QUBO SVM trained with ideal simulations (suffix "i")
    \item QUBO SVM  trained with ideal simulations, with average voting ensemble strategy (suffixes "i" and "opt")
    \item QUBO SVM trained with 1000 noisy simulations (suffixes "N" and the number 1000)
    \item QUBO SVM trained with 1000 noisy simulations, with average voting ensemble strategy (suffixes "N" and "opt")
    \item QUBO SVM trained with 500 noisy simulations (suffixes "N" and 500)
    \item QUBO SVM trained with 500 noisy simulations, with average voting ensemble strategy (suffixes "N", "opt" and 500)
    \item QUBO SVM trained with 100 noisy simulations (suffixes "N" and 100)
    \item QUBO SVM trained with 100 noisy simulations, with average voting ensemble strategy (suffixes "N", "opt" and 100)
    \item QUBO SVM trained with ideal simulations, used as second layer model in a stacked configuration. Now the classical ML models used in the first layer are: Naive Bayes, Random Forest, Logistic Regression and KNN. (suffixes "i" and "stack")
    \item QUBO SVM trained with noisy simulations (sampled with 500 shots) and used as second layer model in a stacked strategy. The first layer exploits the following classical ML models: Naive Bayes, Random Forest, Logistic Regression and KNN. (suffixes "N" and "stack")
\end{itemize}
For completeness, the QUBO SVM models used in this work exploit linear kernel.

\end{document}